\documentstyle[12pt]{article}
\def\sq{\hbox {\rlap{$\sqcap$}$\sqcup$}}
\def\sq{\hbox {\rlap{$\sqcap$}$\sqcup$}}
\def\R{ {\rm R \kern -.31cm I \kern .15cm}}
\def\C{ {\rm C \kern -.15cm \vrule width.5pt \kern .12cm}}
\def\Z{ {\rm Z \kern -.27cm \angle \kern .02cm}}
\def\N{ {\rm N \kern -.26cm \vrule width.4pt \kern .10cm}}
\def\1{{\rm 1\mskip-4.5mu l} }
\def\lsim{\raise0.3ex\hbox{$<$\kern-0.75em\raise-1.1ex\hbox{$\sim$}}}
\def\gsim{\raise0.3ex\hbox{$>$\kern-0.75em\raise-1.1ex\hbox{$\sim$}}}
\def\noi{\noindent}

\def\beq{\begin{equation}}   \def\eeq{\end{equation}}
\def\bea{\begin{eqnarray}}  \def\eea{\end{eqnarray}}
\def\nn{\nonumber}
\def\noi{\noindent}
\def\beeq{\begin{eqnarray}} \def\eeeq{\end{eqnarray}}
\newcommand\mysection{\setcounter{equation}{0}\section}
\renewcommand{\theequation}{\thesection.\arabic{equation}}
\newcounter{hran} \renewcommand{\thehran}{\thesection.\arabic{hran}}

\def\bmini{\setcounter{hran}{\value{equation}}
   \refstepcounter{hran}\setcounter{equation}{0}
   \renewcommand{\theequation}{\thehran\alph{equation}}\begin{eqnarray}}

\def\bminiG#1{\setcounter{hran}{\value{equation}}
\refstepcounter{hran}\setcounter{equation}{-1}
\renewcommand{\theequation}{\thehran\alph{equation}}
\refstepcounter{equation}\label{#1}\begin{eqnarray}}

\def\emini{\end{eqnarray}\relax\setcounter{equation}{\value{hran}}\renewcommand{\theequation}{\thesection.\arabic{equation}}}

\begin{document}
\centerline{\Large\bf Confinement and Mass Gap in Abelian Gauge}
\vskip 1 truecm

\begin{center}
{\bf Ulrich Ellwanger}$^{\rm a}$\footnote{E-mail :
Ulrich.Ellwanger@th.u-psud.fr}, {\bf Nicolas Wschebor}$^{\rm a,\, 
b}$\footnote{E-mail :
Nicolas.Wschebor@th.u-psud.fr}\par \vskip 5 truemm

   $^{\rm a}$ Laboratoire de Physique Th\'eorique\footnote{Unit\'e
Mixte de Recherche - CNRS - UMR 8627}\par
  Universit\'e de Paris XI, B\^atiment
210, F-91405 Orsay Cedex, France\par \vskip 5 truemm

  $^{\rm b}$ Institutos de F{\'\i}sica\par
Facultad de Ciencias (Calle Igu\'a 4225, esq. Mataojo)\par
and Facultad de Ingenier{\'\i}a (C.C. 30, CP 1100), Montevideo, Uruguay
\end{center}
\vskip 2 truecm

\begin{abstract}
First, we present a simple confining abelian pure gauge theory.
Classically, its kinetic term is not positive definite, and it contains
a simple UV regularized $F^4$ interaction. This provoques the formation
of a condensate $\phi \sim F^2$ such that, at the saddle point
$\widehat{\phi}$ of the effective potential, the wave function
normalization constant of the abelian gauge fields 
$Z_{eff}(\widehat{\phi})$ vanishes exactly. Then we study $SU(2)$ pure
Yang-Mills theory in an abelian gauge and introduce an auxiliary field
$\rho$ for a BRST invariant condensate of dimension 2, which renders
the charged sector massive. Under simple assumptions its effective low
energy theory reduces to the confining abelian model discussed before,
and the vev of $\rho$ is seen to scale correctly with the
renormalization point. Under these assumptions, the confinement
condition $Z_{eff} = 0$ also holds for the massive charged sector,
which suppresses the couplings of the charged fields to the abelian
gauge bosons in the infrared regime. 
\end{abstract}

\vskip 3 truecm
\noi LPT Orsay 02-72 \par
\noi September 2002 \par

\newpage
\pagestyle{plain}
\baselineskip 18pt

\mysection{Introduction}
\hspace*{\parindent} Various aspects of the confining phase of 
Yang-Mills theories become more transparent in the abelian gauge
\cite{1r,2r}, notably the phenomenon of  monopole condensation
\cite{1r,3r} according to which the vacuum behaves as a dual 
superconductor. Since these monopoles are essentially configurations of
gauge fields belonging to  the $U(1)$ Heisenberg sub-algebras of
$SU(N)$, the abelian subsector of non-abelian gauge  theories plays the
dominant role for this mechanism responsible for confinement. This 
phenomenon is called abelian dominance \cite{4r,5r}. \par

In the abelian gauge the dynamics of the ``abelian'' gauge fields is 
thus expected to differ considerably from the dynamics of the
``charged'' gauge fields  (associated to off-diagonal generators, and
charged with respect to at least one $U(1)$  subgroup). Whereas the
abelian gauge fields are expected to reproduce essentially the
phenomenon of  monopole condensation of compact QED in the confining
phase \cite{6r}, the charged gauge fields are  expected to be massive
and contribute only sub-dominantly to large distance phenomena. This 
massive behaviour of the charged gauge field propagators has been
observed in lattice studies  \cite{5r}. \par

As dynamical origin of the masses of the charged gauge fields 
ghost-antighost condensates \cite{7r,8r} and bi-ghost condensates
\cite{9r} of dimension 2 have been proposed. Notably a particular 
combination of ghost-antighost and gauge field condensates is BRST
invariant (up to a total derivative) both in the abelian gauge and a
generalization of the Lorentz gauge \cite{10r}. If this particular 
condensate is realized, it describes simultaneously the dimension 2
gauge field condensate  discussed independently in \cite{11r,12r} in
the Landau gauge. Note, however, that  the ghost-antighost condensates
in \cite{7r} do not allow for such a BRST invariant extension and
induce thus necessarily a spontaneous breakdown of BRST symmetry.\par

In [7-9] the formation of the ghost condensate has been related to  the
presence of four ghost interactions in the corresponding gauges. From
the Nambu-Jona-Lasinio  model \cite{13r} it is well known that
four Fermi interactions can provoke the formation of bilinear
condensates. However, here the coefficient of the four ghost
interaction is  proportional to an a priori arbitrary gauge parameter
$\alpha$. Hence the scale of the condensate  is not given by the
confinement scale $\Lambda_{QCD}$ (we continue to denote this scale  by
an index QCD, although we will consider only pure Yang-Mills theories),
unless one  fine-tunes $\alpha$ to be proportional to the first
coefficient of the $\beta$ function [7-9]. One of the purposes of the
present paper is to present a different mechanism for the formation  of
a dimension 2 condensate, which leads automatically to its
proportionality to $\Lambda_{QCD}^2$. \par

Moreover, one would like to learn more about the relation between the 
dimension~2 condensate and the properties of the confining phase as the
area law of the Wilson loop, the condensation of monopoles, and the
vanishing of the effective wave function normalisation constants
$Z_{eff}$ \cite{14r,15r}. (Here $Z_{eff}$ corresponds to $Z_3^{-1}$ in
\cite{15r}, and the relation between the vanishing of $Z_{eff}$ and the
Kugo-Ojima criterion for confinement \cite{16r} has been discussed in
\cite{17r}.) In the Landau gauge, relations of the dimension 2
condensate with confinement have been discussed in \cite{11r,18r}.\par

The description of monopole condensation requires either the
introduction of the t'Hooft monopole  operator \cite{19r} or the
introduction of an antisymmetric tensor field $B_{\mu\nu}$  [20-23],
which is dual to a monopole condensate and couples to the surface of
the Wilson loop. In \cite{22r} the relation between monopole
condensation, the area  law of the Wilson loop and $Z_{eff} = 0$ has
been discussed in a formulation of the Yang-Mills  partition function
involving $B_{\mu\nu}$, and in \cite{24r} these relations have been 
shown to hold in a solvable abelian model in the large $N$ limit.\par

In the present paper we will not introduce a $B_{\mu\nu}$ field, and 
concentrate on $Z_{eff}=0$ as a criterium for confinement. In the first
part of the paper (chapter 2) we present a simple confining abelian
gauge theory, which involves a non-renormalizable interaction $\sim
\lambda^2 F^4$ and has to be equipped with a UV cutoff as, e.g., in the
form of a decreasing momentum dependent form factor in the interaction
term. Also the kinetic term is assumed to show some non-trivial
momentum dependence. (Both these features of the abelian model are
obtained in chapter 3, where the abelian model is derived from $SU(2)$
pure Yang-Mills theory in the abelian gauge.)

Then we will introduce a dimension 4 condensate $\phi$ for the (abelian)
field strength squared. We show that the effective potential for this
condensate can develop a saddle point, which corresponds exactly to
$Z_{eff}(\phi) = 0$ and where the propagator of the abelian gauge
fields behaves like $q^{-4}$ for $q^2 \to 0$. As in the  model in
\cite{24r} this saddle point is only ``visible'' if one introduces an
infrared  cutoff, and studies the limit where the infared cutoff is
removed. We will introduce a  momentum space cutoff $k^2$;
alternatively the system can be placed into a finite volume, and then
the infinite volume limit can be considered. In this limit the
confining  saddle point turns into an essential singularity of the
effective potential which, however, remains finite at this point. 

In chapter 3 we turn to $SU(2)$ pure Yang-Mills theory in the abelian
gauge, and introduce an auxiliary field $\rho$ for the above-mentioned
condensate of dimension 2, which renders the charged gauge fields (and
ghosts) massive. After integrating out these charged fields, the
remaining effective action for the abelian gauge field (and $\rho$)
ressembles to the abelian model of chapter 2. Repeating the steps
of chapter 2 one finds that the ``confining'' saddle point of the
effective potential now also fixes $\rho$ to be of ${\cal
O}(\Lambda_{QCD}^2)$. 

We will argue that $Z_{eff} = 0$ for the abelian gauge  fields induces
also $Z_{eff} = 0$ for the charged gauge fields (and ghosts), invoking 
renormalization group arguments (as in \cite{14r,15r}). This has less
dramatic effects on the (massive) propagators of the charged fields,
but now the couplings of the charged fields to the abelian gauge
fields, which are induced by the $U(1)$ covariant derivatives in the
kinetic terms of the charged fields, vanish in the infrared.\par

Interestingly, the essential features of the mechanism for confinement
considered here are visible already in a loop expansion of the
effective action, once the corresponding auxiliary fields are
introduced, and once certain  perturbatively irrelevant terms in the
effective action are taken into account. \par

In chapter 4 we conclude, summarizing the essential properties of 
our approach.

\mysection{A Confining Abelian Gauge Theory}
\hspace*{\parindent} A class of confining abelian gauge theories has
been discussed in refs. \cite{24r}. These models involve antisymmetric
tensor fields $B_{\mu\nu}$, and are solvable in a $1/N$ limit. In the
present chapter we present a simplified version of these models: first,
we do not introduce antisymmetric
tensor fields $B_{\mu\nu}$ and second, we confine ourselves to $N=1$.
The field content is thus just an abelian gauge field $A_\mu$. In the
absence of a $1/N$ limit the ``solution'' of the model is no longer
quantitatively exact, but its qualitative features remain the same (see
the discussion below).

In the presence of antisymmetric tensor fields the area law of the
Wilson loop is easily obtained in the confining phase, since
antisymmetric tensor fields couple naturally to the enclosed surface. In
the formulation with abelian gauge fields only, the criterium for
confinement becomes a $q^{-4}$ behaviour of its propagator in the
infrared limit, which implies a vanishing wave function renormalization
constant $Z_{eff}$ as in \cite{14r,15r}. (The relation between a
$q^{-4}$ behaviour of the gauge field propagator and the area law of the
Wilson loop has been discussed in \cite{25r}.) 

The simplest confining abelian gauge model involves just a kinetic term
including higher derivatives, $\frac{1}{4} F_{\mu\nu}  Z^A(-\sq)
F_{\mu\nu}$, and a $\lambda^2 F^4$ interaction. For the model to be
confining, $Z^A$ and the dimensional coupling $\lambda^2$ have to
satisfy some inequality (see below), notably $Z^A(0)$ has to be
negative. Clearly this model is ``non-renormalizable'', and has to be
supplemented with an UV cutoff $\Lambda$. This makes sense, since it is
only believed to correspond to an ``effective low energy theory'' of a
non-abelian gauge theory in the abelian gauge, where the off-diagonal
gauge fields are massive.

We will implement an UV cutoff by supplementing the $\lambda^2 F^4$
interaction with a momentum dependent form factor, which decreases
sufficiently rapidly at large momenta. Again this form of the UV cutoff
is motivated by the idea that the $\lambda^2 F^4$ interaction is induced
by loops of massive non-abelian gauge fields, hence the UV cutoff is
naturally of the order of the non-abelian gauge field masses. Actually,
in the Yang-Mills case the corresponding decay of the induced form
factor is not sufficiently rapid in order to prevent logarithmic
divergences, which require the standard counter terms of Yang-Mills
theories. Since we are interested in the infrared behaviour of the
present model, however, we will simplify the treatment of its UV
behaviour and replace the ``soft'' UV cutoff by a``sharp'' UV cutoff.

Thus we take as action of the model (including a standard gauge fixing
term)

\beq
\label{2.1e}
S(A_\mu) = \int d^4 x\left \{ \frac{1}{4} F_{\mu\nu} Z^A(-\sq)
F_{\mu\nu} + \frac{\lambda^2}{8} {\cal O}(x) {\cal O}(x) + \frac {1}{2
\beta} (\partial_\mu A_\mu)^2 \right \}
\eeq

\noi with

\beq
\label{2.2e}
{\cal O}(x) = \int {\cal D}q\ e^{iqx} \int {\cal D}p\ \theta (\Lambda^2
- p^2)\ F_{\mu\nu}(p+q) F_{\mu\nu}(q-p)
\eeq

\noi where ${\cal D}q \equiv d^4q/(2\pi)^4$, and $F_{\mu\nu} =
\partial_\mu A_\nu - \partial_\nu A_\mu$ or its Fourier transform. The
$\theta$-function introduced in (\ref{2.2e}) suffices to regularize all
UV divergences in the approximation considered below. For $Z^A(q^2)$ we
make the choice

\beq
\label{2.3e}
Z^A(q^2) = Z^A_0 + \frac{a_1 q^2}{a_2 \Lambda^2 + q^2}
\eeq

\noi with $Z^A_0 + a_1 > 0$ such that $Z^A(q^2 \to \infty) > 0$,
but later we will allow for $Z^A(0) = Z^A_0 < 0$. Again this choice will
be motivated in the next chapter by the idea that $S(A_\mu)$ in
(\ref{2.1e}) corresponds to an effective low energy theory. The
constants $a_1$ and $a_2$ are assumed to be positive and of ${\cal
O}(1)$.

Next we rewrite the (Euclidean) partition function of the model,
introducing an auxiliary field $\phi(x)$ for the operator (\ref{2.2e}):
 
\beq
\label{2.4e}
e^{-G(J)} = 
\int{\cal D}A{\cal D}\phi e^{-\int d^4 x \left \{\frac{1}{4}
F_{\mu\nu} Z^A(-\sq) F_{\mu\nu} - \frac{1}{8} \phi^2(x) +
\frac{\lambda}{4} \phi(x){\cal O}(x) + \frac{1}{2 \beta} (\partial_\mu
A_\mu)^2 - J_\mu A_\mu \right \}}
\eeq

The coefficient of the $\phi^2$ term in the exponent in (\ref{2.4e})
seems to have the ``wrong'' sign. However, the Gaussian path integral
over $\phi(x)$ is still well defined by analytic continuation and gives
back the original action (\ref{2.1e}); corresponding procedures for
auxiliary fields with ``wrong'' sign quadratic terms are well known from,
e.g., supersymmetric theories in the formulation with auxiliary fileds
$F$ and $D$.

Now the $A_\mu$ path integral is Gaussian; the terms quadratic in
$A_\mu$ can be written as (up to the gauge fixing term, and for
constant $\phi$ for simplicity)

\beq
\label{2.5e}
\frac{1}{4} \int {\cal D}q F_{\mu\nu}(-q) Z^A_{eff}(\phi, q^2)
F_{\mu\nu} (q)
\eeq

\noi with

\beq
\label{2.6e}
Z^A_{eff}(\phi, q^2) = Z^A_0 + \frac{a_1 q^2}{a_2 \Lambda^2 + q^2} +
\lambda \phi \theta (\Lambda^2 - q^2)\ \ .
\eeq

As usual we allow ourselves to interchange the $A_\mu$ and $\phi$ path
integrals in (\ref{2.4e}). The logarithm of the determinant of the
Gaussian $A_\mu$ path integral then contributes to the effective
potential $V_{eff}(\phi)$, which has to be used to determine the saddle
point of the remaining $\phi$ path integral. The relevant point is that
the saddle point $\widehat{\phi}$ of $V_{eff}(\phi)$, which represents
the confining phase, will correspond precisely to
$Z^A_{eff}(\widehat{\phi}, 0) = 0$.

The Coleman-Weinberg contribution of the Gaussian $A_\mu$ path integral
to $V_{eff}(\phi)$ reads (in the Landau gauge $\beta = 0$)

\beq
\label{2.7e}
\Delta V(\phi) = \frac{3}{2} \int\frac{q^2dq^2}{16 \pi^2}\ \ln\left (
Z^A_{eff} (\phi, q^2)\right )\ \ .
\eeq

Note that, due to the $\theta$ function in (\ref{2.6e}), all
$\phi$-dependent terms in (\ref{2.7e}) are ultraviolet finite; these
$\phi$-dependent terms remain unchanged by introducing an UV cutoff
$\Lambda^2$ for the $q^2$ integral and omitting the $\theta$ function
in $Z^A_{eff}$.

Since $Z^A_{eff} (\phi, q^2)$ may turn negative for small $q^2$ and
small $\phi$, the infrared behaviour of the $q^2$ integral in
(\ref{2.7e}) is very delicate. Its correct behaviour can only be
obtained by i) introducing an infrared cutoff $k^2$ (for simplicity, we
employ a sharp cutoff of the $q^2$ integral; the final result does not
depend, however, on this choice), ii) study the saddle point(s) of
$V_{eff}(\phi)$ for finite $k^2$ and iii) take the limit $k^2 \to 0$ at
the end.

Hence, instead of (\ref{2.7e}), we write

\beq
\label{2.8e}
\Delta V(\phi) = \frac{3}{2} \int_{k^2}^{\Lambda^2}\frac{q^2dq^2}{16
\pi^2}\ \ln\left ( Z^A_{eff} (\phi, q^2)\right )\ \ 
\eeq

\noi where now the $\theta$ function on the right hand side of
(\ref{2.6e}) is replaced by $1$.

The result of the $q^2$ integral is most easily written in terms of the
combination

\beq
\label{2.9e}
\Sigma (\phi) = \frac{a_2 \Lambda^2 Z^A_{eff}(\phi, 0)}{a_1 +
Z^A_{eff}(\phi, 0)}
\eeq

\noi where

\beq
\label{2.10e}
Z^A_{eff}(\phi, 0) = Z^A_0 + \lambda \phi\ \ .
\eeq

Now the total potential $V(\phi) = - \frac{1}{8}\phi^2 + \Delta V(\phi)$
becomes

\bea
\label{2.11e}
V(\phi) &=& -\frac{1}{8} \phi^2 + \frac{3}{64 \pi^2}\left [ (\Sigma^2 -
k^4)\ \ln(\Sigma + k^2) + (\Lambda^4 - \Sigma^2)\ \ln(\Sigma +
\Lambda^2) \right .
\nn \\  && \left .  
+\  \Sigma (\Lambda^2 - k^2) +(\Lambda^4 - k^4)\ \ln(a_1 +
Z^A_{eff}(\phi, 0))\right ]\nn \\ && + \ (\phi{\rm -independent})\ .
\eea

The saddle point condition then reads

\bea
\label{2.12e}
0 &=& \frac{dV(\phi)}{d\phi}{\Big |}_{\widehat{\phi}} \nn \\ 
&=& - \frac{1}{4} \widehat{\phi} + \frac{3 a_1 a_2 \lambda
\Lambda^2}{32 \pi^2 (a_1 + Z^A_{eff}(\widehat{\phi}, 0))^2}\left [
\Sigma\  \ln\left (\frac{\Sigma + k^2}{\Sigma + \Lambda^2} \right ) +
\Lambda ^2 - k^2\right . \nn \\ &&
\left . +\ \frac{(\Lambda^4-k^4)(a_1+Z^A_{eff}(\widehat{\phi}, 0))} {2
a_1 a_2 \Lambda^2} \right ]
\eea

In the limit $k^2 \to 0$ the product of $\Sigma$ with the logarithm in
(\ref{2.12e}) can show the following subtle behaviour:

\bea  \label{2.13e}
&&k^2 \to 0\ , \nn \\
&&\Sigma \to 0_{-\varepsilon}\ , \nn \\
&&\Sigma \ \ln \left ( {\Sigma + k^2 \over \Sigma + \Lambda^2} 
\right ) \to \ K = {\rm const.}   \eea

\noi where the constant K is positive and can be chosen such that 
(\ref{2.12e}) is satisfied for $k^2 \to 0$, provided

\beq
\label{2.14e}
\frac{1}{4}\widehat{\phi} - \frac{3 a_2 \lambda \Lambda^4}{32 \pi^2 a_1}
\left ( 1+\frac{1}{2 a_2} \right ) > 0
\eeq

\noi which we assume in the following.
Note that $\Sigma \to 0$ corresponds to

\beq
\label{2.15e}
Z^A_{eff}(\widehat{\phi}, 0) = Z^A_0 + \lambda \widehat{\phi} =  0
\eeq

\noi which has already been used in order to derive (\ref{2.14e}) from
(\ref{2.12e}). Note also that the saddle point (\ref{2.13e}) would be
invisible, if we would put $k^2 = 0$ from the beginning: at the
corresponding value $\widehat{\phi}$ (corresponding to $\Sigma = 0$) 
the potential $V(\widehat{\phi}){\Big |}_{k^2 = 0}$ and its first
derivatives are finite, but all higher derivatives diverge. Only after
regularisation of this singularity (through the infrared cutoff $k^2$)
one finds that this essential singularity of $V(\phi)$ contains a
``hidden'' saddle point.

Eq. (\ref{2.15e}) corresponds to the result announced above: at the
confining saddle point (or in the confining phase) the auxiliary field
$\phi$, which corresponds to a condensate $\left < F_{\mu\nu}
F_{\mu\nu}\right >$, arranges itself such that $Z^A_{eff} = 0$ exactly
(without fine tuning).

However, the original parameters of the model have to satisfy some
inequality for the confining phase to exist: from Eqs. (\ref{2.14e}) and
(\ref{2.15e}) one finds easily

\beq
\label{2.16e}
Z^A_0 < -\frac{3 a_2 \lambda^2 \Lambda^4}{8 \pi^2 a_1} \left ( 1 +
\frac{1}{2 a_2}\right )\ ,
\eeq

\noi i.e. notably

\beq
\label{2.17e}
Z^A_0 < 0
\eeq

\noi for $a_1$, $a_2 > 0$, which we do assume.
Eq. (\ref{2.17e}) explains the formation of the condensate $\phi \sim
\left < F_{\mu\nu} F_{\mu\nu} \right >$: now the action (\ref{2.1e}) is
unstable at the origin of constant modes of $F_{\mu\nu}$ already
classically (the classical $A_\mu$ propagator, for $\phi = 0$, would be
Tachyonic for $q^2 \to 0$).

The remarkable point is, however, that the condensate $\widehat{\phi}$
arranges itself in the confining phase such that the $A_\mu$ propagator
in the background $\widehat{\phi}$ shows a $q^{-4}$ behaviour for $q^2
\to 0$ (which is related, of course, to $Z^A_{eff}(\widehat{\phi}) =
0$): after replacing $\phi$ by $\widehat{\phi}$ in the exponent of the
partition function (\ref{2.4e}), i.e. after approximating the $\phi$
path integral by its saddle point, the $A_\mu$ propagator can be
obtained from the inverse of $\frac{1}{2}\delta^2 G(J)/\delta
J_\mu(-q)\delta J_\nu(q)$. In the Landau gauge $\beta \to 0$ one
finds

\beq
\label{2.18e}
P^A_{\mu\nu} = \left ( \delta_{\mu\nu} - \frac{q_\mu q_\nu}{q^2} \right
)  \frac{a_2 \Lambda^2 + q^2}{a_1 q^4}
\eeq

\noi which coincides with the expression for $P^A_{\mu\nu}$ in
confining models with antisymmetric tensor fields $B_{\mu\nu}$
\cite{22r,24r}.

The saddle point approximation for the $\phi$ path integral can be
rendered exact within a $1/N$ expansion \cite{24r}, i.e. after
replacing $A_\mu$ by $A^a_\mu$, $a = 1\dots N$, and rescaling the
coupling correspondingly. In the present case the $\phi$ path integral
is, in principle, not trivial. Note, however, that
$d^2V(\phi)/d\phi^2|_{\widehat{\phi}} = -\infty$, i.e. the $\phi$
propagator vanishes in the confining phase at vanishing momentum. Also,
the coupling of $\phi$ to $F_{\mu\nu} F_{\mu\nu}$ is equipped with an
UV regulator (form factor), hence perturbation theory in powers of this
coupling has a good chance to converge rapidly. A detailed study of
this problem is, however, beyond the scope of the present paper.

\mysection{Mass Gap and Confinement in $SU(2)$ Yang-Mills Theory}
\hspace*{\parindent} As stated in the introduction, we consider pure 
$SU(2)$ Euclidean Yang-Mills theory in a (continuum version of) the
(maximal) abelian  gauge [26-28]. $A_{\mu}$ denotes the abelian gauge
field associated to the $U(1)$ subgroup,  and $W_{\mu}^{\pm}$ the
remaining charged gauge fields. The classical action reads

\beq
\label{3.1e}
S = \int d^4x \left \{ {\cal L}_{YM} + {\cal L}_{GF} \right \}
\eeq

\noi where ${\cal L}_{YM}$ is the Yang-Mills Lagrangian

\bea
\label{3.2e}
{\cal L}_{YM} &=& {1 \over 4} \left ( \partial_{\mu} A_{\nu} - 
\partial_{\nu} A_{\mu} \right )^2 
+ {1 \over 2}
\left ( D_{\mu} W_{\nu}^+ - D_{\nu} W_{\mu}^+ \right ) \left (
D_{\mu} W_{\nu}^- - D_{\nu} W_{\mu}^- \right )\nn \\
&&+ {ig \over 2} \left ( \partial_{\mu} A_{\nu} - \partial_{\nu} 
A_{\mu} \right ) \left (
W_{\mu}^+ W_{\nu}^- - W_{\mu}^- W_{\nu}^+ \right ) 
- {g^2 \over 4} \left ( W_{\mu}^+ W_{\nu}^- - W_{\mu}^- W_{\nu}^+ 
\right )^2 
 \ . \nn \\
\eea

\noi Here $D_{\mu}$ denote the $U(1)$ covariant derivatives 
$\partial_{\mu} \pm ig A_{\mu}$. After elimination of the
Nakanishi-Lautrup auxiliary fields the gauge  fixing part ${\cal
L}_{GF}$ reads

\bea
\label{3.3e}
{\cal L}_{GF} &=& {1 \over 2 \beta} \left (  \partial_{\mu} A_{\mu} 
\right )^2 + {1 \over
\alpha} D_{\mu} W_{\mu}^+ D_{\nu} 
W_{\nu}^- \nn \\
&&+ \partial_{\mu} \bar{c}^3 \left ( \partial_{\mu} c^3 + ig \left ( 
W_{\mu}^+ c^- -
W_{\mu}^- c^+ \right )\right ) + D_{\mu} \bar{c}^+ D_{\mu} c^-\nn \\
&&+ D_{\mu} \bar{c}^- D_{\mu} c^+ + g^2 \left ( W_{\mu}^+ \bar{c}^- - 
W_{\mu}^- \bar{c}^+\right
) \left ( W_{\mu}^+ \bar{c}^- - W_{\mu}^- c^+\right ) \nn \\
&&- \alpha \ g^2\ \bar{c}^+c^- \bar{c}^-c^+ \ . \eea

\noi The neutral ghosts $c^3$, $\bar{c}^3$ actually decouple and will 
play no role in the
following. (There are no vertices involving $c^3$.) \par

Now we introduce an auxiliary field $\rho$  for the bilinear dimension
2 condensate

\beq
\label{3.4e}
W_{\mu}^+ W_{\mu}^- + \alpha \left ( \bar{c}^+ c^- + \bar{c}^- c^+
\right ) \ . \eeq

\noi Under BRST transformations this operator transforms into the 
total derivative
$\partial_{\mu} (W_{\mu}^+c^- + W_{\mu}^- c^+)$ \cite{10r}; for the 
explicit BRST transformations
corresponding to the conventions implicit in ${\cal L}_{GF}$ see 
\cite{28r}. \par

The introduction of $\rho$ corresponds to adding to $S$ 
the complete square

\beq
\label{3.5e}
{\cal L}_\rho = {1 \over {2g^2}} \left ( \rho + 
g^2 W_{\mu}^+ W_{\mu}^- + \alpha g^2
\left (  \bar{c}^+ c^- + \bar{c}^- c^+ \right ) \right )^2 \ , \eeq

\noi and of course  $\rho$ has to transform under BRST  transformations
in the same way as the negative of the operator (\ref{3.4e}). It is
understood that now a $\rho$ path integral has to be performed.\par

Note that, when adding (\ref{3.5e}) to $S$, we made no  effort to
cancel the quartic ghost interaction term in (\ref{3.3e}) as in [7-9];
the powers of $g^2$ in eq. (\ref{3.5e}) have just been introduced in
order to facilitate their bookkeeping. It is of course  true that, once
this term has been cancelled, $S$ is quadratic in the charged  ghosts,
and the ghost path integral can be performed trivially. We do not
believe, however, that  the resulting contribution to the effective
potential of $\rho$ is dominant and  fixes its vev. We will identify
another contribution below, which is more relevant for a small enough
gauge parameter $\alpha$ (but still $\alpha \sim {\cal O}(1)$). In any
case the absence of the four ghost interaction is no scale invariant
statement, since it  is re-generated by $W_{\mu}^{\pm}$-loops (which
are not $1/N$-suppressed, as in solvable  4-Fermi-models). Also the
physical consequences of an auxiliary field as introduced in 
(\ref{3.5e}) should be independent from the conventions chosen for the
corresponding  coefficients; in (\ref{3.5e}) we choose, for simplicity,
conventions such that the induced mass  terms ${\cal L}_{m}$ for the
charged fields are simply expressed in terms of $\rho$:

\beq  \label{3.6e}
{\cal L}_{m} = \rho \ W_{\mu}^+ W_{\mu}^- + \alpha \rho \left ( 
\bar{c}^+c^- + \bar{c}^- c^+
\right ) \ . \eeq

Next we wish to integrate over the charged fields $W_{\mu}^{\pm}$,
${c}^{\pm}$. However, in order to control the UV divergences, one
should integrate simultaneously over  the high momentum modes of the
abelian field $A_{\mu}(p^2)$ with, say, $p^2 > \Lambda^2$. Of  course
it is not trivial to implement such an intermediate scale in a gauge
(or BRST) invariant  way. The best one can do is to implement the
constraint $p^2 > \Lambda^2$ in a Wilsonian sense  (i.e. by modifying
the $A_{\mu}$-propagators correspondingly) which allows to control the 
BRST symmetry with the help of modified Slavnov-Taylor identities
\cite{29r}. For our subsequent  qualitative results and its essential
features the details of this procedure will play no  role, however.
After having renormalized the UV divergences by, e.g., dimensional 
regularization, we are left with an induced effective action
$\Gamma_{eff}(A_{\mu}, \rho )$,  which is a functional of the low
momentum modes of $A_{\mu}$ and of $\rho$. \par

Thus we rewrite the full Yang-Mills path integral -- including the path
integral over $\rho$ -- as

\beq
\label{3.7e}
\int{\cal D}A{\cal D}W{\cal D}c{\cal D}\bar{c}{\cal D}\rho 
e^{-\int d^4x\{{\cal L}_{YM} + {\cal L}_{GF} + {\cal L}_{\rho} \}} 
= \int{\cal D}A_{<\Lambda^2}{\cal D}\rho e^{-\Gamma_{eff}(A,\rho)}
\eeq

\noi where the index $< \Lambda^2$ attached to ${\cal D}A$ denotes the
restriction to modes with $p^2 \lsim \Lambda^2$, and where a $U(1)$
gauge fixing term (the first term in (\ref{3.3e})) is understood in
$\Gamma_{eff}$.

Let us first have a look at the term quadratic in $A_{\mu}$ in
$\Gamma_{eff}(A, \rho)$. Due to  the $U(1)$ gauge invariance it has to
be of the form

\beq
\label{3.8e}
\int {\cal D} q \ {1 \over 4} \ F_{\mu\nu} (-q) \left ( Z_0^A(\rho , 
\mu^2) + f^A(q^2, \rho )
\right ) F_{\mu\nu}(q)\ \ . \eeq

\noi Here we have suppressed the dependence on the gauge parameter 
$\alpha$ which we assume to be of ${\cal O}(1)$ subsequently such that,
from eq. (\ref{3.6e}), the masses of all charged fields are of ${\cal
O}(\rho)$. Of course $Z_0^A(\rho, \mu^2)$ is of the form $Z^A_0(\rho,
\mu^2) = 1 +$ loop  corrections, and we define the splitting between
$Z_0^A(\rho, \mu^2)$ and the $q^2$-dependence parametrized by 
$f^A(q^2,\rho)$ such that $f^A(0, \rho ) = 0$. 

Next we discuss some particular features of the scale anomaly in 
abelian gauges. A natural choice for the running gauge coupling $g_R$
(but not necessarily a  physical one, see below) is the coupling $g$ in
the $U(1)$ covariant derivative $D_{\mu} =  \partial_{\mu} \pm ig
A_{\mu}$, {\it after} rescaling $A_{\mu}$ such that its kinetic term
(\ref{3.8e}) is  properly normalized (at $q^2 = 0$). In the case
(\ref{3.8e}) this immediately leads to

\beq
\label{3.9e}
g_R^2 = {g^2 \over Z_0^A (\rho, \mu^2)}
\eeq

\noi where $g^2$ is constant. The derivative of $g_R^2$ with respect 
to its dimensionful arguments gives the $\beta$ function in a herewith
defined  renormalization scheme. In our case one finds by inspecting
the diagrams which contribute to $Z_0^A$, and  taking into account that
the circulating charged fields have masses given by $\rho$, that to 
one loop order $Z_0^{A(1)}$ is independent of the infrared cutoff
$\Lambda^2$ of the abelian gauge fields, since no internal 
$A_{\mu}$-propagators appear. Thus $Z_0^{A(1)}(\rho , \mu^2)$ depends
on $\rho$ as dictated by the  universal one-loop coefficient $\beta_0$
of the $\beta$ function (cf. \cite{27r}):

\beq
\label{3.10e}
Z_0^{A(1)}(\rho , \mu^2) = 1 - g^2 \beta_0 \ \ell n \left ( {\mu^2 
\over c_1 \rho}\right ) + {\cal O}(g^4) 
\quad \hbox{with}\ \beta_0 = {11 \over 24 \pi^2} \ , \eeq

\noi where $\mu^2$ is the scale where $g^2$ is defined, and $c_1$ is 
an arbitrary coefficient.
\par

Next we consider the $q^2$ dependence of $f ^A(q^2, \rho )$ in 
(\ref{3.8e}). By definition (by choosing the coefficient $c_1$ in
(\ref{3.10e}) correspondingly) it  vanishes at $q^2 = 0$, and for large
$q^2 \gg \rho$ the same scale anomaly arguments force it to behave, to
one loop order, as

\beq
\label{3.11e}
f^A(q^2 , \rho ) \sim g^2 \beta_0 \ \ell n \left ( {q^2 \over \rho} 
\right ) \ .
\eeq

\noi For our subsequent purposes it will be sufficient to replace the 
logarithmic rise in (\ref{3.11e}) by a positive constant for $q^2 \to
\infty$, since momenta with $q^2 \gg \rho$ will be cutoff anyhow (see
below). Thus we parametrize $f^A$ as

  \beq
\label{3.12e}
f^A(q^2 , \rho ) \sim {a_1q^2 \over a_2\rho + q^2} \ ,
\eeq

\noi with $a_1$, $a_2$ positive numerical coefficients of ${\cal
O}(g^2)$, ${\cal O}(1)$, respectively. Now the kinetic terms in
(\ref{3.8e}) are of the form of the kinetic terms of the model in
chapter 2, provided we identify $Z^A_0$ in (\ref{2.3e}) with
$Z^A_0(\rho, \mu^2)$ in (\ref{3.8e}) (or (\ref{3.10e})), and
$\Lambda^2$ in the denominator in (\ref{2.3e}) with $\rho$ in the
denominator in (\ref{3.12e}).

Next we will discuss the leading perturbatively irrelevant terms in
$\Gamma_{eff}(A_\mu, \rho)$ generated by loops of the (massive) charged
fields. Again, by $U(1)$ invariance (and a discrete  $Z_2$-symmetry
$A_{\mu} \to - A_{\mu}$), these have to be quartic in the abelian
field  strength $\sim (F_{\mu\nu})^4$, convoluted with a form factor of
the four in- or out-going momenta.  Dimensional analysis dictates that,
for large and equal Euclidean momenta $q^2$, the form  factor has to
decay like $q^{-4}$. For distinct momenta this form factor will be a
complicated  function including possible open Lorentz indices.\par

For our subsequent purposes it will be sufficient to assume the 
presence of a particularly simple structure among all possible terms
$\sim (F_{\mu\nu})^4$,  which we parametrize as

\bminiG{3.13e}
\label{3.13ae}
{\lambda^2 \over 8} \int d^4x \ {\cal O}(x) \ {\cal O}(x) \ ,
\eeeq
\beeq
  \label{3.13be}
{\cal O}(x) = \int {\cal D} q\ e^{iqx} \int {\cal D}p \ 
F_{\mu\nu}(p+q)\ h(p^2) \
F_{\mu\nu}(q-p) \emini

\noi where, for the above reasons, $h(p^2)$ has to decay like  $p^{-2}$
for large $p^2$. $\lambda^2$ is of the order of

\beq
\label{3.14e}
\lambda^2 = \frac{{\widehat{\lambda}}^2g^4}{16 \pi^2 \rho^2}
\eeq

\noi where $\widehat{\lambda}$ is of ${\cal O}(1)$. 

Note that the expression (\ref{3.13ae}) appears with a positive  sign.
This follows from the limit of large field strengths $F_{\mu\nu}$ (at
vanishing  momenta), where the dependence of the induced
effective action on $F_{\mu\nu}^2$ must be of the form 

\beq  \label{3.15e}
\int d^4 x \ F_{\mu\nu}^2 \left ( 1 + g^2\beta_0 \ \ell n \left 
( {C_1 + F_{\mu\nu}^2 \over
C_2} \right ) \right ) \ , \eeq

\noi for some constants $C_1 \sim C_2 \sim  \Lambda_{QCD}^4$, in order
to reproduce the scale anomaly \cite{30r}. Expanding (\ref{3.15e}) to
${\cal  O}(F_{\mu\nu}^4)$ gives a positive coefficient. 

Subsequently we replace $h(p^2)$ by a ``sharp'' cutoff,

\beq
\label{3.16e}
h(p^2) \sim \theta(\rho - p^2)\ .
\eeq

Note that the scale of the ``UV cutoff'' in (\ref{3.16e}) has to be of
${\cal O}(\rho)$, since the contribution (\ref{3.13ae}) to 
$\Gamma_{eff}(A_\mu, \rho)$ was generated by loops of the charged
fields with masses of ${\cal O}(\rho)$ and consequently  $h(p^2)$
decays only for $p^2 \gg \rho$.

With the sharp cutoff we throw away logarithmic effective 2-loop
divergences, which would contribute to the $A_\mu$ propagator (and
hence to the renormalization of $g^2$) once 1-loop diagrams with the
effective vertex (\ref{3.13e}) are computed. Here, however, we are not
interested in the 2-loop $\beta$-function, but in capturing the
essential features of the infrared regime.

Hence, in the present approximation, $\Gamma_{eff}(A_\mu, \rho)$
coincides with $S(A_\mu)$ of the previous model in (\ref{2.1e}),
provided we perform the following replacements of the parameters of the
model in chapter 2:

\noi i) replace $Z^A(q^2)$ in (\ref{2.1e}), (\ref{2.3e}) by
$Z^A(q^2, \rho)$, with

\beq
\label{3.17e}
Z^A(q^2, \rho) = Z^A_0(\rho, \mu^2) + \frac{a_1 q^2}{a_2 \rho + q^2}
\eeq

\noi where, to one loop order, $Z^A_0(\rho, \mu^2)$ has the form given
in Eq. (\ref{3.10e});

\noi ii) replace $\Lambda^2$ in the $\theta$ function in (\ref{2.6e}) by
$\rho$, and $\lambda$ by a $\rho$-dependent expression of the form
(\ref{3.14e}).

At this point it may be helpful to summarize the procedure and the
approximations, under which the Yang-Mills theory turns into the
confining abelian model of section 2:

1) An auxiliary field $\rho$ for a dimension 2 condensate is
introduced, such that all charged gauge fields and ghosts are massive
for $\left < \rho \right > \neq 0$ (which remains to be shown).

2) The path integral over the charged gauge fields and ghosts, as
well as over the "high momentum modes"  (with $p^2 > \Lambda^2$) of the
neutral gauge field $A_\mu$, is performed.

3) The resulting effective action $\Gamma_{eff}(A_\mu, \rho)$ is
not computed exactly, but assumed to be well approximated by the
following form:

a) The term quadratic in $A_\mu$ in $\Gamma_{eff}$ is given by the
wave function normalization function $Z^A(q^2, \rho)$ of eq.
(\ref{3.17e}), whose dependence on $q^2$ and $\rho$ is known (from the
scale anomaly) for large $q^2$ or $\rho$, but parametrized by an
"educated guess" for small $q^2$ and $\rho$: For vanishing $q^2$,
$Z^A(0, \rho) \equiv Z^A_0(\rho, \mu^2)$ is assumed to be given by an
extrapolation of the one loop (or scale anomaly) result (\ref{3.10e}),
valid for large $\rho$, towards small $\rho$. Then, notably, $Z^A(0,
\rho)$ turns negative for $\rho < {\cal O}(\Lambda^2_{QCD})$. This (and
only this) assumption is crucial for the subsequent results.

The dependence of $Z^A(q^2, \rho)$ on $q^2$ has to interpolated between
$Z^A(0, \rho)$ and the known scale anomaly result (\ref{3.11e}) for
large $q^2$. Our subsequent qualitative results do not depend on the
precise dependence of $Z^A(q^2, \rho)$ on $q^2$; therefore, in order to
allow for an analytic computation of the integral over $q^2$ appearing
below, we parametrize its $q^2$ dependence by the simple analytic
structure (\ref{3.17e}), which replaces the logarithmic rise for $q^2
\to \infty$ by a constant $a_1$.

b) The term quartic in $A_\mu$ in $\Gamma_{eff}$ is approximated by
the $F^4$ term described in eqs. (\ref{3.13e}) -- (\ref{3.16e}), i.e.
more complicated tensorial structures are dropped and the form factor
is replaced by the sharp cutoff (\ref{3.16e}). As in the case of the
simplified parametrization of the $q^2$ dependence of $Z^A(q^2, \rho)$
above, these approximations do not affect qualitatively the results
below, but allow for subsequent analytic computations. Finally, terms of
higher order in $A_\mu$ are dropped in $\Gamma_{eff}$, again for
computational simplicity.

The Yang-Mills partition function (\ref{3.7e}) can then be rewritten, as
in the previous model, invoking an auxiliary field $\phi$. Including a
source for $A_\mu$ as in (\ref{2.4e}) one obtains

\beq
\label{3.18e}
e^{-G(J)} = \int{\cal D}A_{<\Lambda^2}{\cal D}\rho{\cal D}\phi e^{-\int
{\cal D} q \left \{\frac{1}{4} F_{\mu\nu} Z^A_{eff} (\phi,q^2,\rho)
F_{\mu\nu} + \frac{1}{2 g^2} \rho^2 + \widehat{V}(\rho) - \frac{1}{8} 
\phi^2
+ \frac{1}{2 \beta} (q_\mu A_\mu)^2 - J_\mu A_\mu \right \}}
\eeq

\noi with

\beq
\label{3.19e}
Z^A_{eff} (\phi,q^2,\rho) = Z^A (q^2, \rho) + \lambda \phi \theta (\rho
- q^2)\ \ .
\eeq

The term ${\rho^2}/{2g^2}$ in the exponent in (\ref{3.18e})
originates from ${\cal L}_\rho$ (\ref{3.5e}), and $\widehat{V}(\rho)$
from the path integrals over $W^{\pm}_\mu$ and $c^\pm$. In refs.
[7--10] one loop expressions for $\widehat{V}(\rho)$ have been used in
order to fix the vev $\widehat{\rho}$ of $\rho$ (the saddle point of the
$\rho$ path integral), with the unsatisfactory result that
$\widehat{\rho}$ depends correctly on $\Lambda_{QCD}$ only for a
fine-tuned value of the gauge parameter $\alpha$. Here we argue,
instead, that $\widehat{\rho}$ is determined by another contribution to
$V(\rho)$, which is obtained in analogy to the model in chapter~2. Then
$\widehat{\rho}$ depends automatically on $\Lambda_{QCD}$ as it should
(see below).

The full effective potential $V_{eff}(\phi, \rho)$ is obtained by
performing the $A_\mu$ path integral in (\ref{3.18e}), i.e.

\beq
\label{3.20e}
V_{eff}(\phi, \rho) = \frac{1}{2g^2} \rho^2 + \widehat{V}(\rho)
-\frac{1}{8} \phi^2 + \Delta V(\phi, \rho)
\eeq

\noi with $\Delta V(\phi, \rho)$ as in Eq. (\ref{2.8e}):

\beq
\label{3.21e}
\Delta V(\phi, \rho) = \frac{3}{2} \int^{\Lambda^2}_{k^2} \frac{q^2
dq^2}{16 \pi^2}\ \ln\left ( Z^A_{eff} (\phi, q^2, \rho)\right )
\eeq

If we replace the upper limit $\Lambda^2$ of the $q^2$ integral in
(\ref{3.21e}) by $\rho$, the result for $\Delta V(\phi, \rho)$ can be
obtained from the previous results in chapter 2 after simple
substitutions. The ``error'' is then given by a $q^2$ integral ranging
from $\rho$ to $\Lambda^2$. However, below we are only interested in
$\Delta V(\phi, \rho)$ near the saddle point $\widehat{\rho}$; choosing,
at the end, $\Lambda^2 \sim \widehat{\rho}$, this error can be made
arbitrarily small.

The $\phi$-dependent terms in $\Delta V(\phi, \rho)$ are then given by
$V(\phi)$ as in Eq. (\ref{2.11e}), after replacing $\Lambda^2$ by
$\rho$ everywhere, and $Z^A_{eff}(\phi, 0)$ by

\beq
\label{3.22e}
Z^A_{eff}(\phi, 0,\rho) = Z^A_0(\rho, \mu^2) + \lambda \phi\ \ .
\eeq

The $\phi$-independent terms in $\Delta V(\phi, \rho)$, neglected in
(\ref{2.11e}), are quadratic in $\rho$ and of ${\cal O}(\rho^2/16
\pi^2)$, hence negligible compared to the ``tree level'' term ${\rho^2}
/{2g^2}$ in (\ref{3.20e}).

In order to determine the extrema $\widehat{\phi}$, $\widehat{\rho}$ of
$V_{eff}(\phi, \rho)$ we first look for extrema with respect to $\phi$,
plug the resulting expression $\widehat{\phi}(\rho)$ back into
$V_{eff}(\widehat{\phi}(\rho), \rho)$, and minimize with respect to
$\rho$ at the end.

The equation for extrema with respect to $\phi$ can again be taken from
chapter 2, Eq. (\ref{2.12e}), with the substitutions above. Hence we
obtain the following important results:

\noi i) again a confining saddle point exists, which shows the behaviour
(\ref{2.13e}) and, in analogy with (\ref{2.15e}),

\beq
\label{3.23e}
Z^A_{eff}(\widehat{\phi}, 0,\rho) = Z^A_0(\rho, \mu^2) + \lambda
\widehat{\phi} = 0\ \ .
\eeq

Eq. (\ref{3.23e}) fixes the dependence of $\widehat{\phi}(\rho)$ on
$\rho$:

\beq
\label{3.24e}
\widehat{\phi}(\rho) = - \lambda^{-1} Z^A_0 (\rho, \mu^2)
\eeq

\noi (Recall that here $\lambda$ depends on $\rho$, cf. Eq.
(\ref{3.14e}).)

\noi ii) the necessary condition for the confining phase to exist, the
analog of Eq. (\ref{2.16e}) with $\lambda^2$ as in Eq. (\ref{3.14e}),
now reads

\beq
\label{3.25e}
Z^A_0(\rho, \mu^2) < -{\cal O}\left ( \left (\frac{g^2}{8 \pi^2}\right
)^2 \right )\ \ ,
\eeq

\noi i.e. essentially

\beq
\label{3.26e}
Z^A_0 (\rho, \mu^2) < 0\ \ .
\eeq

It remains to show that a saddle point $\widehat{\rho}$ with the above
properties actually exists. Neglecting in $V_{eff}$ terms of ${\cal
O}(\rho^2/16 \pi^2)$ relative to $\rho^2/2g^2$ as above, $\Delta
V_{eff}(\widehat{\phi} (\rho), \rho)$ can be dropped and
$V_{eff}(\widehat{\phi} (\rho), \rho)$ is simply given by

\bea
\label{3.27e}
V_{eff}(\widehat{\phi}(\rho), \rho) &=&  \frac{1}{2g^2} \rho^2 -
\frac{1}{8}  \widehat{\phi}(\rho)^2 + \widehat{V} (\rho)\nn \\
&=& \frac{1}{2g^2} \rho^2 - 
\frac{2 \pi^2 \rho^2}{\widehat{\lambda}^2g^4} 
\left ( Z^A_0 (\rho, \mu^2) \right )^2 + \widehat{V}(\rho)
\eea

\noi where we have used Eqs. (\ref{3.24e}) and (\ref{3.14e}). Let us
first neglect $\widehat{V}(\rho)$ in (\ref{3.27e}), which was generated
by the path integral over the charged fields: Using the one loop
expression (\ref{3.10e}) for $Z^A_0 (\rho, \mu^2)$ one finds that
$V_{eff}(\widehat{\phi}(\rho), \rho)$  vanishes for $\rho \to 0$, and
is negative for small $\rho$, since $Z_0^{A (1)}$ diverges
logarithmically for $\rho \to 0$. Then there appears a minimum, where
$Z^{A(1)}_0 (\rho, \mu^2)$ is negative and of ${\cal O}(g^2)$. This
minimum constitutes the desired confining saddle point
$\widehat{\rho}$. If we continue to increase $\rho$, 
$V_{eff}(\widehat{\phi}(\rho), \rho)$ increases until it reaches a
maximum where $Z^A_0 (\rho, \mu^2)$ is positive (still of ${\cal
O}(g^2)$), and for $\rho \to \infty$ it is unbounded from below as in
the case of $V(\phi)$ for $\phi \to \infty$. Hence all we have to
require from $\widehat{V}(\rho)$ is that it does not {\it destroy} the
desired saddle point, i.e. that it is small enough for the
corresponding value $\widehat{\rho}$. 

$\widehat{V}(\rho)$, as
computed in refs. [7--10], is proportional to the arbitrary gauge
parameter $\alpha$, since it has its origin in the quartic ghost
interaction term (the last term in eq. (\ref{3.3e})): With the
definition (\ref{3.5e}) for $\rho$ we have $\widehat{V}(\rho) \sim
(\alpha/16 \pi^2) \rho^2 \ln \rho^2$. Since $\alpha$ is
multiplicatively renormalized \cite{28r}, it is always possible to chose
gauges (bare values of $\alpha < {\cal O}(1)$, since $\widehat{\lambda}$
in (\ref{3.27e}) is of ${\cal O}(1)$, and $Z^A_0(\widehat{\rho})$ is of
${\cal O}(g^2)$) such that the effect of 
$\widehat{V}(\rho)$ in (\ref{3.27e}) is negligible. The above result is
thus valid in abelian gauges with $\alpha$ smaller than some critical
value of ${\cal O}(1)$.

From

\beq
\label{3.28e}
Z^A_0(\widehat{\rho}, \mu^2) \sim -\ {\cal O}(g^2)
\eeq

\noi one easily obtains, using the 1-loop expression (\ref{3.10e}) for 
$Z^A_0(\widehat{\rho}, \mu^2)$,

\beq  \label{3.29e}
\widehat{\rho} \sim c_1^{-1} \ \mu^2 \ e^{-{1 \over {g^2\beta_0}} (1 + 
{\cal O}(g^2))} \sim \Lambda^2_{QCD}
\eeq

\noi which is the result announced above and which has to be
contrasted with the results in [7-9], where the  dimension 2 condensate
scales correctly only for a particular choice of the gauge parameter
$\alpha$, $\alpha = {1 \over 2} \beta_0$.

Actually, if one computes $Z^A_0(\widehat{\rho}, \mu^2)$ to higher loop
orders, Eq. (\ref{3.28e}) (or a more accurate minimization of the
effective potential with respect to $\rho$) always {\underbar
{defines}} a scale $\Lambda^2_{QCD}$ through its solution
$\widehat{\rho}$. From Eq. (\ref{3.24e}), with $\lambda \sim
\rho^{-1}$, one finds immediately that also
$\widehat{\phi}(\widehat{\rho}) \sim \Lambda^2_{QCD}$ and hence

\beq
\label{3.30e}
\left < F_{\mu\nu} F_{\mu\nu} \right > \sim
\lambda^{-1}\widehat{\phi} \sim \Lambda^4_{QCD}\ \ .
\eeq

Clearly, in terms of the running coupling $g_R^2(\rho )$ defined in 
(\ref{3.9e}) through $Z_0^A(\widehat{\rho}, \mu^2)$, the condition
(\ref{3.25e}) for the confining saddle point corresponds to 
$g_R^2(\widehat\rho ) < 0$, i.e. $g_R^2$ has ``passed'' a Landau
singularity. However, the kinetic term of the $A_\mu$ field in the full
effective action in the presence of the condensate $\widehat{\phi}$ is
proportional to $Z_{eff}(\widehat{\phi}, q^2, \widehat{\rho})$ (cf. Eq.
(\ref{3.19e})) and is never negative, but vanishes for $q^2 \to 0$.
Hence, if one insists on defining a running coupling $g(q^2)$ in terms
of $Z^A_{eff}(\widehat{\phi}, q^2, \widehat{\rho})^{-1}$, it diverges
in the confining phase for $q^2 \to 0$. We do not find this convention
very appropriate, however, since it refers to the virtualities $q^2$ of
the abelian gauge fields $A_\mu$ only. We prefer to continue to
parametrize the vertices by the constant $g^2$, keeping the $q^{-4}$
behaviour of the propagator $P^A_{\mu\nu}(q^2)$.

Let us discuss the couplings of the charged gauge fields to $A_\mu$ in
more detail. The $U(1)$ gauge symmetry (left unbroken up to the
standard $U(1)$ gauge  fixing term) allows to separate these couplings
into two classes: i) couplings involving  the abelian field strength
$F_{\mu\nu}$. Here additional derivative(s) act on $A_{\mu}$, which 
soften infrared divergencies of loops involving the $A_{\mu}$
propagator emerging  from such vertices. ii) couplings involving the
$U(1)$ covariant derivative $D_{\mu}=  \partial_{\mu} \pm ig A_{\mu}$
acting on charged fields. These appear in the $U(1)$ invariant  kinetic
energy terms for $W_{\mu}^{\pm}$ and the charged ghosts in the
Lagrangian  (\ref{3.2e}), (\ref{3.3e}). In the quantum effective action
these kinetic terms appear multiplied with  wave function
renormalization constants $Z_W$, $Z_c$, respectively. In the  following
we will study the behaviour of these constants in the infrared limit
and find that they  vanish; this suppresses automatically the couplings
of the neutral to charged gauge fields in the infrared. 

For a most general parametrization of the quantum effective action  the
wave function renormalization constants should actually be replaced by
functions of  the (covariant) Laplacian $D_{\mu} D_{\mu}$ or, in
momentum space, by functions of $q^2$ plus  the corresponding couplings
to the neutral gauge field $A_{\mu}$ required by $U(1)$ gauge
invariance. In general, for large Euclidean non-exceptional momenta
$q^2 \to \infty$, the parameters of the  quantum effective action
approach their ``bare'' values ($Z_W$, $Z_c = 1$). Subsequently we 
replace $Z_W$ and $Z_c$ by constants for simplicity, i.e. we compute
these functions at $q^2 =  0$. Hence their vanishing does not suppress
the associated couplings to $A_{\mu}$ completely, just the associated
form factors in the limit $q^2 \to 0$. \par

In the simplified parametrization of the quantum effective action  with
constant $Z_W$, $Z_c$, the relevant terms (quadratic in $W_{\mu}^{\pm}$
and the charged ghosts) read
 
\bea
\label{3.31e} 
&&{Z_W \over 2} \left ( D_{\mu} W_{\nu}^+ - D_{\nu} W_{\mu}^-\right ) 
\left ( D_{\mu} W_{\nu}^- -
D_{\nu} W_{\mu}^-\right ) + \rho W_{\mu}^+ W_{\mu}^- \nn \\
&&+ {Z_W \over \alpha} \left ( D_{\mu} W_{\mu}^+\right ) \left (
D_{\nu}  W_{\nu}^- \right ) + Z_c \left ( D_{\mu} \bar{c}^+ D_{\mu} c^-
+ D_{\mu} \bar{c}^- D_{\mu} c^+  \right ) \nn \\
&&+ \alpha \rho \left ( \bar{c}^+c^- + \bar{c}^- c^+ \right ) \ .
\eea

\noi Here we have included the mass terms originating from ${\cal 
L}_m$ in (\ref{3.6e}). To one loop order $Z_W$ and $Z_c$ get
renormalized only by ``rainbow'' diagrams where the rainbow corresponds
to a $A_{\mu}$ propagator of the form (\ref{2.18e}) (we continue to
work in the Landau gauge $\beta \to 0$ for the  abelian sector). For
convenience we introduce the notation $\varphi = \{W_{\mu}^{\pm} ,
{c}^{\pm}\}$ in the following. Our aim is now to derive a
renormalization group  equation for $Z_{\varphi}$. From the $A_{\mu} -
\varphi - \varphi $ vertices from (\ref{3.31e}) one finds that the one
loop contributions to $Z_{\varphi}$ due to the rainbow diagrams are

\beq  \label{3.32e}
\Delta Z_{\varphi} = c_{\varphi} \ g^2 \ Z_{\varphi}^2 \int {q^2 dq^2 
\over 16 \pi^2} \
P_{\varphi}(q^2) \ P_A(q^2) \ . \eeq

\noi Here $P_{\varphi}$ are the massive  $W_{\mu}^{\pm}/ c^{\pm}$
propagators,

\beq  \label{3.33e}
P_{\varphi}(q^2) = {1 \over Z_{\varphi} q^2 + (\alpha) \widehat{\rho}}
\eeq

\noi (where the factor $\alpha$ appears only for the ghosts, and equals
1 for $W_\mu^\pm$), and  $P_A(q^2)$ reads, from (\ref{2.18e}),

\beq  \label{3.34e}
P_A(q^2) = {a_2 \widehat{\rho} + q^2 \over a_1 q^4} \ .
\eeq

\noi The constants $c_{\varphi}$ in (\ref{3.32e}) read (for $SU(2)$)

\bea  \label{3.35e}
&&c_{\varphi = W} = - {1 \over 6} (17 - 3 \alpha ) \ , \nn \\
&&c_{\varphi = c} = - 3 \ .
\eea

\noi Note that, for $\alpha$ not too large, we have $c_{\varphi} < 0$.
\par

In order to derive a renormalization group equation from eq. 
(\ref{3.32e}) we proceed as in the case of the computation of the
effective potential $V(\rho , \phi  )$: We introduce an infrared cutoff
$k^2$ for the $q^2$ integral in (\ref{3.32e}), and replace the
constants $Z_\phi$ by $Z_\phi(k^2)$. Then we take the derivatives with
respect to $k^2$  of both sides of eq. (\ref{3.32e}) and study the
running of $Z_{\varphi} (k^2)$. As before  such an infrared cutoff
could be implemented in the ``Wilsonian'' way by modifying the 
propagator $P_A(q^2)$, but for the present purposes it is sufficient to
simply cutoff the $q^2$  integral in (\ref{3.32e}) at its lower end.
(Also this procedure can be re-interpreted as a  ``sharp'' Wilsonian
cutoff function in $P_A(q^2)$). After introduction of this infrared
cutoff  $k^2$, and taking the derivative $d/dk^2$ on both sides of eq.
(\ref{3.32e}), one obtains

\beq  \label{3.36e}
k^2 \ {dZ_{\varphi}(k^2) \over dk^2} = - Z_{\varphi}^2(k^2)\ 
{c_{\varphi}g^2 \over 16 \pi^2} \
k^4 \ P_{\varphi} (k^2) \ P_A(k^2) \eeq

\noi which becomes in the deep infrared regime $k^2 \ll \widehat{\rho}$
 
\beq  \label{3.37e}
k^2 \ {dZ_{\varphi}(k^2) \over dk^2} \cong - Z_{\varphi}^2(k^2)\ 
{c_{\varphi}g^2 \over 16 \pi^2}
\ {a_2 \over (\alpha )} \ .
\eeq

\noi (Again the factor $\alpha$ in the denominator appears only for 
the charged ghosts). Note that, if $P_A(q^2)$ would not behave as
$q^{-4}$ for $q^2 \to 0$, $Z_\varphi(k^2)$ would stop to run with $k^2$
for $k^2 \ll \widehat\rho$. Eq. (\ref{3.37e}) is easily solved with the
result

\beq  \label{3.38e}
Z_{\varphi}(k^2) = {{Z_{\varphi}(\Lambda^2)} \over {1 + 
Z_{\varphi}(\Lambda^2) {c_{\varphi}g^2 \over 16 \pi^2}\ {a_2 \over
(\alpha )}\ \ln \left ( {k^2 \over \Lambda^2} \right )}}
\eeq

\noi and hence, for $k^2 \to 0$ and with $c_{\varphi} < 0$, we obtain 
$Z_{\varphi} (0) = 0$ as announced. One can check that the
contributions from  multi-rainbow-diagrams to the running of
$Z_{\varphi}(k^2)$ are suppressed by higher powers of the bare 
coupling $g^2$. Also, the renormalization of the $\rho$ vertex to the
charged fields (or their mass terms) is infrared finite precisely because
of the massiveness of the charged fields, in spite of the $q^{-4}$
behaviour of the $A_\mu$ propagator.
\par

The question arises, however, whether this suppression of the  $A_{\mu}
- \varphi - \varphi$ couplings for $q^2 \to 0$ does not invalidate the
contributions of  the $\varphi$ loops to the effective action
$\Gamma_{eff}(A_{\mu}, \rho )$, which have been used  extensively
before. The essential features of these contributions, on the other
hand, arise  either from virtualities $q^é$ of the $\varphi$-fields
$(W_{\mu}^{\pm},  {c}^{\pm}$) which are very large $(q^2 \to \infty$),
or from $q^2 \sim \rho$ where the massiveness  (infrared finiteness) of
the charged propagators is used. These features remain valid even if
the  $A_{\mu} - \varphi - \varphi$ couplings become suppressed for $q^2
\ll \rho$, and  $Z_{\varphi}$ in the $\varphi$ propagators becomes
replaced by $Z_{\varphi}(q^2)$ with  $Z_{\varphi}(0) = 0$.\par

This result completes the confinement criterion $Z_{eff} = 0$ of 
\cite{14r,15r}, which now holds for all fields of pure Yang-Mills
theory including the charged ones: The simple poles of their
propagators disappear, since their mass terms remain finite. In the
abelian  gauge the $A_{\mu} - W_{\mu}^+ - W_{\mu}^-$ vertex form factor
will of course not be  symmetric in the 3 external momenta; the present
result applies to the limit of vanishing  momenta squared of the
charged fields $W_{\mu}^{\pm}$. Nevertheless this behaviour of the
vertex  form factor helps to suppress infrared divergencies of higher
order loop diagrams, which helps to render the present approach stable
with respect to higher loop orders.

\mysection{Conclusions and Outlook}
\hspace*{\parindent} The aim of the present paper is to show how the 
cooperation of two condensates of dimension 2 and 4, respectively,
generates both the  confinement condition $Z_{eff} = 0$ and a mass gap
for the charged fields in the abelian  gauge in the confining phase.
The study of a subtle saddle point of the effective potential  is
required to this end, which is only visible in the limit where an
artificial infrared  cutoff goes to zero. \par

The only role of the dimension 2 condensate is actually to give masses
to all charged fields, i.e. the charged gauge fields $W_{\mu}^{\pm}$
and the charged ghosts. We choose the BRST invariant combination of
bilinear charged gauge fields and ghosts here; the ghost condensates in
refs. [7, 9] are claimed to induce masses for the charged gauge fields
by loops and could, in principle, do the same job. The formal proof of
gauge invariance of the Yang-Mills quantum effective action relies,
however, on the vanishing of the expectation values of all BRST-exact
operators. This is no longer guaranteed, if the BRST symmetry is
spontaneously broken.

Here we have concentrated on the condition $Z_{eff} = 0$ for 
confinement, which corresponds to the absence of coloured asymptotic
states. It would be quite  straightforward, however, to introduce
additional (auxiliary) antisymmetric tensor fields  $B_{\mu\nu}$ for
the abelian field strength, and to study the corresponding effective
action. As in the  case of the $1/N$-solvable abelian models \cite{24r}
this would make the relation with monopole  condensation and the area
law for the Wilson loop explicit. \par

Also for simplicity we have insisted on simple parametrizations of  the
$q^2$ and $\rho$ dependences of various terms in the effective action,
in order to allow for an analytic study of the appearance of the
confining saddle point. It would not be too  hard to compute these
dependencies exactly (to one loop order); then, however, the confining
saddle point  induced by $A_{\mu}$ loops could be studied only
numerically and would be somewhat less obvious. \par

On the other hand our parametrizations reproduce the essential 
features of the relevant terms in the effective action, which allows to
study the essential  mechanism behind the confining saddle point:
Without the condensate $\phi \sim F_{\mu\nu}^2$, 
$Z_{eff}^A(\widehat{\rho})$ would turn negative for $\widehat{\rho}$
small enough. This  corresponds to a non-convexity of
$\Gamma_{eff}(F_{\mu\nu}^2)$ around the origin, which is impossible.
The condensate $\phi$  then renders $\Gamma_{eff}(F_{\mu\nu}^2)$
semi-convex, which corresponds to the non-analytic behaviour of its
effective potential. \par

The fact that these essential features are visible already after the 
computation of one loop diagrams should not make one believe that
confinement is  ``perturbative'': If one eliminates all auxiliary
fields by its equations of motion at the very end it  becomes clear
that, by computing $\Gamma_{eff}$ in its presence, one has implicitly
summed  up an infinite number of loops.\par

Nevertheless the question arises whether the present approach would 
allow for quantitatively stable higher order corrections, once lowest
orders are computed with  sufficient precision. (Given that even
perturbation theory is not asymptotically stable,  this is evidently a
rather ambitious program.) More concretely, this corresponds to the
question  whether possible infrared divergencies from higher order
corrections can be controlled  or, better, shown to be absent. Two
steps in this direction are, in the present approach, i)  the
massiveness of the charged gauge fields (and ghosts), and ii) the
vanishing of the wave  function renormalization constants of the
charged fields in the infrared. Notably this latter phenomenon will
suppress very long range correlation functions between operators
involving charged fields in spite of the $q^{-4}$  behaviour of the
abelian propagator.\par

Finally we remark that the simultaneous presence of a mass gap (of the
charged fields) and confining interactions (as induced by the
abelian sector) can most likely be made explicit only in the abelian
gauge. This gauge evidently plays an essential  role in the present
approach, which describes a quite explicit dynamical mechanism behind 
the confining phase in continuum Yang-Mills theory. An interesting task
for the future will be the study of the constraints on the full
effective action $\Gamma_{eff}$ which arise from the Slavnov-Taylor
identities in the abelian gauge (suitably generalized due to the
presence of the auxiliary fields), once the present results on some
selected terms in $\Gamma_{eff}$ are taken into account.

\end{document}